\documentclass[
5p,
times,sort&compress
]
{elsarticle}

\usepackage{graphicx}
\usepackage{xcolor}
\usepackage[colorlinks=true,linkcolor=blue,urlcolor=blue,citecolor=blue,bookmarks=true,bookmarksopenlevel=2]{hyperref}
\usepackage{amsmath}

\usepackage{microtype}

\journal{Journal of Quantitative Spectroscopy \& Radiative Transfer}

\newcommand{\alert}[1]%
{%
\textcolor{red}{#1}
}%

\newcommand{\ket}[1]{\left\vert#1\right\rangle}

\begin{document}

\begin{frontmatter}
\title{MCDF-RCI predictions for structure and width of $K\alpha_{1,2}$ x-ray line of Al and Si}
\author {Karol Kozio{\l}} 
\ead{mail@karol-koziol.net}
\address{Faculty of Chemistry, Nicolaus Copernicus University, 87-100 Toru\'n, Poland}

\begin{abstract}
Multiconfiguration Dirac-Fock and Relativistic Configuration Interaction methods have been employed to predict the structure and the width of $K\alpha_{1,2}$ x-ray lines of Al and Si. The influences of electron correlation and inclusion of possible satellite contributions on spectra structure have been studied. 
The widths of $K$ and $L_{2,3}$ atomic levels of Al and Si have been also computed. 
\end{abstract}


\end{frontmatter}

\section{Introduction}

X-ray emission spectroscopy (XES) has, for many years, provided valuable information about the electronic structure of atoms and molecules. In particular, the width of spectral line can provide valuable information about the mechanism and dynamics of hole states’ creation in atoms induced by photon, electron, or ion beams (see e.g. \cite{lifetimes-prl,qstate,hci2012-ir} and references therein). 

In the work of Polasik et al. \cite{lifetimes-prl} there was demonstrated that a major part of $K^h\alpha_{1,2}$ linewidths for atoms with 20$\le$Z$\le$30 can be explained by using the Open-shell Valence Configuration (OVC) model. 
In this paper the width of $K\alpha_{1,2}$ x-ray lines of Al and Si have been predicted by using the OVC approach, followed by Multiconfiguration Dirac-Fock (MCDF) calculations supported by Relativistic Configuration Interaction (RCI) calculations. 
The $K\alpha_{1,2}$ x-ray line refers to the $1s^{-1}\to2p^{-1}$ transitions, where the notation $1s^{-1}$ refers to a hole in the $1s$ shell, and $2p^{-1}$ refers to a hole in the $2p$ shell. 
The theoretical predictions for $K\alpha_{1,2}$ linewidth have been followed by computation of $K$ and $L_{2,3}$ atomic level width by using a combination of MCDF and multiconfiguration Dirac-Hartree-Slater (DHS) methods.

It is worth underlining that RCI calculations are performed mostly for high-Z and/or multiple ionized atoms \cite{cff2,cff3,hao2010}, and the RCI method was used for theoretical consideration of x-ray spectra which originated from one-hole states (so-called diagram lines) in only a few cases \cite{chantler2009,chantler2010,lowe,chantler-Ka-Ti,chantler-Cu-sat,chantler-grant,lowe-abinitio}. 
Then, it is interesting to test the RCI approach in the case of diagram x-ray lines for low-Z atoms, where correlation effects are not overwhelmed by relativistic effects.

\section{Theoretical background}

\subsection{Relativistic calculations}

The methodology of MCDF calculations performed in the present studies is similar to that which has previously been published in a number of papers (see, e.g., \cite{gr6,grant2}).
The effective Hamiltonian for an N-electron system is expressed by
\begin{equation}
H = \sum_{i=1}^{N} h_{D}(i) + \sum_{j>i=1}^{N} C_{ij},
\end{equation}
where $h_{D}(i)$ is the Dirac operator  for $i$-th electron, and the terms $C_{ij}$ account for electron-electron interactions. 
The latter is the sum of the  Coulomb interaction operator and the transverse Breit operator.
An atomic state function (ASF) with the total angular momentum $J$ and parity $p$ is assumed in the form
\begin{equation}
\Psi_{s} (J^{p} ) = \sum_{m} c_{m} (s) \Phi ( \gamma_{m} J^{p} ),
\end{equation}
where $\Phi ( \gamma_{m} J^{p} )$ are configuration state functions (CSFs), $c_{m} (s)$ are the configuration mixing coefficients for state $s$, and $\gamma_{m}$ represents all information  required  to uniquely define a certain CSF. 

The accuracy of the wavefunction depends on the CSFs included in its expansion \cite{lowe,cff2}. 
Accuracy can be improved by extending the CSF set by including the CSFs originated by excitations from orbitals occupied in the reference CSFs to unfilled orbitals of the active orbital set (i.e. CSFs for virtual excited states). 
This approach is called Configuration Interaction (CI) or, for relativistic Dirac-Fock calculations, Relativistic CI. 
The CI method makes it possible to include the major part of the electron correlation contribution to the energy of the atomic levels. 
The most important thing with the CI approach is to choose a proper basis of CSFs for the virtual excited states. 
It is reached by systematic building of CSF sequences by extending Active Space of orbitals and monitoring concurrently the convergence of self-consistent calculations \cite{cff2,cff3}.

The calculations of radiative transition rates were carried out by means of \textsc{Grasp2k} code \cite{grasp2k} by using the MCDF approach. 
Apart from the transverse Breit interaction, two types of quantum electrodynamics (QED) corrections (self-energy and vacuum polarization) have been included in the perturbational treatment. 
Despite the fact that the choice of model of estimation QED energy contributions in many-electron atoms may be important for high-Z atoms \cite{lowe-se-approx}, it is less important for considered in this work low-Z atoms. 
The initial and the final states of $K\alpha_{1,2}$ transitions were computed separately and the biorthonormal transformation \cite{biorto1,biorto2} was used before calculation of the radiative transition rates. 
The radiative transition rates were calculated in both Coulomb (velocity) \cite{grant1} and Babushkin (length) \cite{cech-bab} gauges. 
The calculations of non-radiative transition rates were carried out by means of \textsc{Fac} code \cite{fac1,fac2} by using the DHS approach. 
The multiconfiguration DHS method, in general, is similar to the MCDF method, but a simplified expression for electronic exchange integrals is used \cite{fac2}.

\subsection{Lifetime of excited states, width of corresponding atomic levels, and fluorescence yields\label{sec:teor-lif-wid}}

Each excited state can be linked to the mean lifetime~$\tau$. The mean lifetime can be defined as a time after which the number of excited states of atoms decreases $e$ times. The mean lifetime is determined by the total transition rate of de-excitation (radiative and non-radiative) processes $W_i$: 
\begin{equation}
\tau = \left(\sum_i W_i\right)^{-1} = \left(\sum_i X_i + \sum_j A_j\right)^{-1}\;,
\label{eq:tau}
\end{equation} 
where $X_i$ is the transition rate of the radiative process and $A_j$ is the transition rate of the non-radiative Auger process. De-excitation processes happen for all possible ways leading to lower energetic states allowed by selection rules.

Due to the energy-time uncertainty principle ($\Delta E \Delta t = \hbar$), the lifetime of excited state $\tau$ is connected to the width of corresponding atomic level $\Gamma$ (note that for one atomic level there may be more than one corresponding excited state) by the relationship 
\begin{equation}
\Gamma = \frac{\hbar}{\tau} = \hbar W = \hbar \sum_i W_i\;. 
\label{eq:gam}
\end{equation}
The natural width of an atomic level can be obtained as a sum of radiative width $\Gamma^{Rad}$ and non-radiative width $\Gamma^{Nrad}$:
\begin{equation}
\Gamma =  \Gamma^{Rad} + \Gamma^{Nrad}\;.
\label{eq:gamma}
\end{equation} 
The relevant yields are linked with these terms, i.e. 
\begin{subequations}
\begin{eqnarray}
&\omega = \frac{\Gamma_X}{\Gamma} = \frac{\sum_i X_i}{\sum_i X_i + \sum_j A_j}\;,&\\
&a = \frac{\Gamma_A}{\Gamma} = \frac{\sum_j A_j}{\sum_i X_i + \sum_j A_j}\;,&
\label{eq:omega}
\end{eqnarray}
\end{subequations}
where $\omega$ is the fluorescence yield and $a$ is the Auger yield.

For open-shell atomic systems for each atomic hole level $(nlj)^{-\alpha}$ (where $\alpha$= 1, 2, \ldots, $2j+1$) there are linked a~lot of hole states $(nlj)_J{}^{-\alpha}$. Therefore, for $i$-th hole state all transition rates $W_{ij}$ corresponding to $j$-th de-excitation process should be considered. 
The radiative part of the natural width $\Gamma^{Rad}_i$ of $i$-th hole state can be determined by using the transition rate of radiative processes $X_{ij}$ according to the formula 
\begin{equation}
\Gamma^{Rad}_i = \hbar \sum_j X_{ij} \;.
\label{eq:grad}
\end{equation}
The radiative part of the natural width of $(nlj)^{-\alpha}$ hole atomic level is taken as the arithmetic mean of the radiative parts to the natural width of the each hole state $(nlj)_J{}^{-\alpha}$, i.e. according to the formula 
\begin{equation}
\Gamma^{Rad} = \frac{\sum_i \Gamma^{Rad}_i}{n} \;,
\label{eq:ave}
\end{equation}
where $\Gamma^{Rad}$ is a radiative part of the natural level width, and $n$ is the number of the hole state corresponding to a given hole level. 

Similarly, the non-radiative part of the natural width can be determined by calculating the transition rates for the non-radiative Auger $A_{ij}$ and Coster-Kronig $C_{ik}$ processes according to the formula 
\begin{equation}
\Gamma^{Nrad} = \frac{\hbar \sum_{i,j} A_{ij}}{n} \;,
\label{eq:gnrad}
\end{equation}
where the designations are analogous to the above. 
Then the total natural width of the atomic hole levels can be determined according to the Eq.~\eqref{eq:gamma}.

Equations \eqref{eq:grad}, \eqref{eq:ave}, and \eqref{eq:gnrad} can be rewritten in the forms:
\begin{equation}
\Gamma^{Nrad} = \hbar \sum_{j} \bar X_{j} \;,
\label{eq:grada}
\end{equation}
\begin{equation}
\Gamma^{Nrad} = \hbar \sum_{j} \bar A_{j} \;,
\label{eq:gnrada}
\end{equation}
where $\bar X_j = \frac{1}{n} \sum_{i} X_{ij}$ is a mean value of transition rates per one $(nlj)_J{}^{-\alpha}$ hole state for $j$-th de-excitation channel, and similarly  for $\bar A$.

\begin{figure}[htb!]
\centering
\includegraphics[width=\columnwidth]{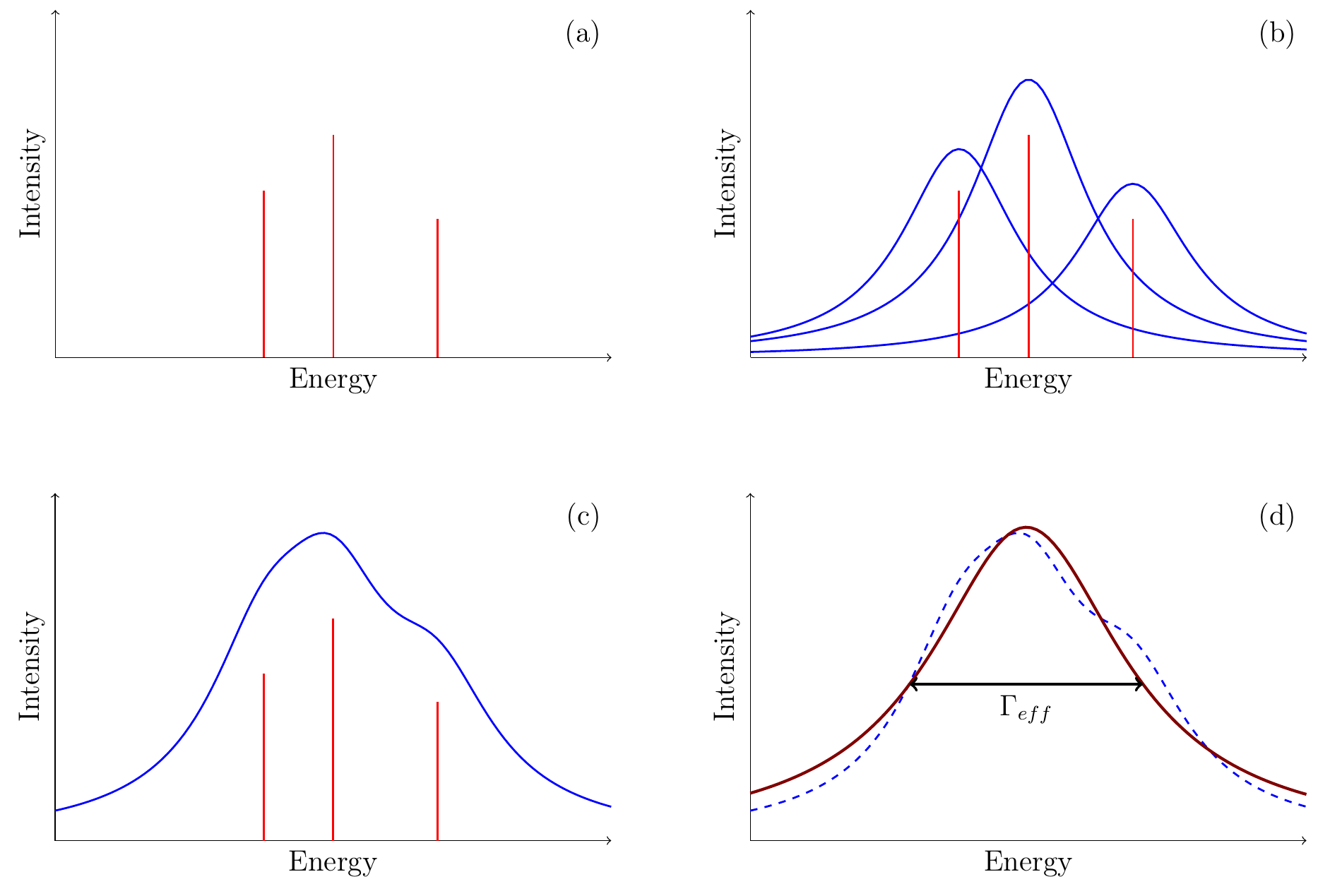}
\caption{\label{eff-wid-proc-idea}Schematic diagram of determination of effective spectral linewidth: (a)~stick spectrum, (b)~Lorentz profile for each transition, (c)~synthetic spectrum as a sum of Lorentz profiles for each transition, (d)~one Lorentz profile fitted to synthetic spectrum.}
\end{figure}

\begin{table*}[!htb]
\caption{\label{tab:kwid}Mean values of transition rates for $K$-shell de-excitation processes and the widths and fluorescence yields of $K$ level for Al and Si. The 'C' and 'B' symbols indicate values calculated by using Coulomb and Babushkin gauges, respectively.}
\begin{tabular*}{\linewidth}{@{\extracolsep{\fill}}ccccccccccccc}
\hline
\multicolumn{13}{c}{~}\\[-1.5ex]
&& \multicolumn{7}{c}{$\bar W_i$ [s$^{-1}$]} &&&&\\[0.5ex]
\cline{3-9}\\[-2ex]
& Number of & \multicolumn{2}{c}{$K\alpha_{1,2}$ [$\times10^{13}$]} & \multicolumn{2}{c}{$K\beta_{1,3}$ [$\times10^{12}$]} & K-LL & K-LM & K-MM & \multicolumn{2}{c}{$\Gamma_K$ [eV]} & \multicolumn{2}{c}{$\omega_K$}\\[0.5ex]
\cline{3-4}\cline{5-6}\cline{10-11}\cline{12-13}\\[-2ex]
& $1s^{-1}$ levels & C & B & C & B & [$\times10^{14}$] & [$\times10^{13}$] & [$\times10^{11}$] & C & B & C & B\\[0.5ex]\hline\\[-1.5ex]
Al & 4 & 2.419 & 2.592 & 0.313 & 0.314 & 5.360 & 2.853 & 4.905 & 0.388 & 0.389 & 0.042 & 0.044\\
Si & 8 & 3.457 & 3.678 & 1.080 & 1.059 & 5.803 & 4.378 & 8.708 & 0.435 & 0.436 & 0.054 & 0.057\\
\hline
\end{tabular*}
\end{table*}

\begin{table*}[!htb]
\caption{\label{tab:lwid}Mean values of transition rates for $L_{2,3}$-shell de-excitation processes, the widths and fluorescence yields of $L_{2,3}$ level, and the widths of $K\alpha_{1,2}$ transitions for Al and Si. The 'C' and 'B' symbols indicate values calculated by using Coulomb and Babushkin gauges, respectively.}
\begin{tabular*}{\linewidth}{@{\extracolsep{\fill}}ccccccccccc}
\hline
\multicolumn{11}{c}{~}\\[-1.5ex]
&& \multicolumn{3}{c}{$\bar W_i$ [s$^{-1}$]} &&&& \\[0.5ex]
\cline{3-5}\\[-2ex]
& Number of & \multicolumn{2}{c}{$L\eta$+$Ll$ [$\times10^{9}$]} & L-MM & \multicolumn{2}{c}{$\Gamma_{L_{2,3}}$ [eV]} & \multicolumn{2}{c}{$\omega_{L_{2,3}}$ [$\times10^{-4}$]} & \multicolumn{2}{c}{$\Gamma_{K\alpha_{1,2}}$ [eV]}\\[0.5ex]
\cline{3-4}\cline{6-7}\cline{8-9}\cline{10-11}\\[-2ex]
& $2p^{-1}$ levels & C & B & [$\times10^{13}$] & C & B & C & B & C & B\\[0.5ex]\hline\\[-1.5ex]
Al & 10 & 5.447 & 5.509 & 3.647 & 0.024 & 0.024 & 1.49 & 1.51 & 0.412 & 0.413\\
Si & 21 & 8.243 & 5.374 & 4.618 & 0.030 & 0.030 & 1.78 & 1.16 & 0.465 & 0.467\\
\hline
\end{tabular*}
\end{table*}

\subsection{Evaluation of effective linewidth}

The width of the $K\alpha_{1,2}$ transition is commonly expressed as:
\begin{equation}
\Gamma_{K\alpha_{1,2}} = \Gamma_{K} + \Gamma_{L_{2,3}}
\label{eq:ka}
\end{equation}
Unfortunately, the measured width of the $K\alpha_{1,2}$ line for low- and medium-Z atoms (when the width of $K\alpha_{1,2}$ transition is comparable to energetic differences within $K^{-1}$ or $L_{2,3}^{-1}$ level sets) is often larger than the width of the $K\alpha_{1,2}$ transition defined in Eq.~\eqref{eq:ka} (see e.g. \cite{sorum,kuchler}). So, there is a need for a more detailed theoretical study. 

The OVC effect mentioned in the Introduction is related to the fact that, in the case of open-shell atoms, there are many initial and final states for each x-ray line. 
The x-ray line consists then of numerous overlapping components having slightly different energies and widths. 
As a consequence of the OVC effect, the effective natural $K\alpha_{1,2}$ linewidths are much larger than
those predicted by Eq.~\eqref{eq:ka}. 

The idea of the OVC effective linewidth evaluation procedure is presented on Fig.~\ref{eff-wid-proc-idea}. 
At first, the stick spectrum (consisting MCDF-calculated transition energy and intensity) for all $K\alpha_{1,2}$ transitions were generated (Fig.~\ref{eff-wid-proc-idea}a). 
Next, the Lorentz profile for each transition were built (Fig.~\ref{eff-wid-proc-idea}b). 
As a width of Lorentz profile the present calculated $K\alpha_{1,2}$ transition width was taken. 
The synthetic spectrum of the whole $K\alpha_{1,2}$ line is given as a sum of Lorentz profiles for each transition (Fig.~\ref{eff-wid-proc-idea}c).
Finally, in order to estimate the effective $K\alpha_{1,2}$ linewidth, the synthetic spectrum was fitted with one Lorentz profile (Fig.~\ref{eff-wid-proc-idea}d). 
It is considered in this work on x-ray spectra energy range that the instrumental broadening of the spectrum (represented in x-ray modelling as a width of Gaussian profile) is close to 0.1 eV \cite{graeffe} or even much smaller \cite{citrin}. Therefore, for simplification, in this study only Lorentz profile fitting is used. 
The fitting process has been performed by using the OriginPro code \cite{originpro}.

It is worth mentioning that fitting a spectrum with only one Lorentz profile is not the best way of fitting in the case of a strongly asymmetric peak, such as an Al or Si $K\alpha_{1,2}$ line. 
For asymmetric x-ray peaks, for example a $K\alpha_{1,2}$ or $K\beta_{1,3}$ lines for low- and medium-Z elements, the best way, characterized by a small $\chi^2_r$ coefficient, is to fit a measured spectrum with a couple of Voigt profiles (or just Lorenzian profiles, if instrumental broadening, given as a Gaussian width, is small). 
This way is established from many years in fitting of experimental spectra of 3p-elements \cite{al-decomposition,graeffe} and 3d-elements \cite{holzer,chantler2006,chantler-Kb-Ti,chantler-Ka-Ti,smale-Kb-V,illig-Ka-Cu}. The higher number of Voigt/Lorenzian profiles used, the better fitting and the smaller $\chi^2_r$ coefficient obtained \cite{illig-Ka-Cu}. 
However, the higher number of profiles used in fitting an experimental spectrum, the bigger the interpretation problems. Are these fitted profiles of physical or only mathematical meaning? This question is particularly important in the case where the obtained fit profiles differ significantly in width because of linking between level\slash{}line width and decay rates of hole state (see Eqs. \ref{eq:tau}, \ref{eq:gam}, and \ref{eq:ka}). Drawing reliable conclusions from such profiles requires a careful understanding of the various x-ray processes and experimental setup \cite{illig-Ka-Cu}. 
The fitting of the $K\alpha_{1,2}$ spectrum with two peaks -- one for $K\alpha_{1}$ line and one for $K\alpha_{2}$ -- seems to be simple and natural. 
However, this way is not justified from a theoretical point of view in the case of low-Z elements (in opposite to medium- and high-Z elements). 
That is because for low-Z elements there is a small $2p_{1/2}$-$2p_{3/2}$ orbital energy splitting and dominance of the LS coupling scheme in the intermediate coupling, so CSFs consisting of $2p_{1/2}$ and $2p_{3/2}$ hole are well mixed in the final states of $K\alpha_{1,2}$ transitions. 

In this work the synthetic spectra have been created from scratch, basing on MCDF calculations, and compared to experimental data using linewidth, given as a width of one Lorentz profile fitted, as a key parameter.

\section{Results and discussion}

\subsection{Width of $K\alpha_{1,2}$ transition}

The mean values of transition rates for $K$- and $L_{2,3}$-shell de-excitation processes, the widths and fluorescence yields of $K$ and $L_{2,3}$ level, and the widths of $K\alpha_{1,2}$ transitions (utilizing both Coulomb's and Babushkin’s calculated values for radiative width) for Al and Si are presented in Tables \ref{tab:kwid} and \ref{tab:lwid}. 
It is worth noticing that because the total widths of $K$ and $L_{2,3}$ levels for low-$Z$ Al and Si are more determined by non-radiative parts, there is only a minor difference between values of the widths of $K$ and $L_{2,3}$ levels and widths of $K\alpha_{1,2}$ transitions obtained from Coulomb's and Babushkin’s calculations. 

The presented values of $K$ level widths, 0.388\slash 0.389 and 0.435\slash 0.436 eV for Al and Si respectively (for Coulomb\slash Babush\-kin gauge respectively), are slightly larger than the recommended values from Campbell and Papp’s work \cite{papp}, i.e. 0.37 eV for Al and 0.43 eV for Si, but smaller than values presented in Krause and Oliver’s work \cite{krause}, i.e. 0.42 eV for Al and 0.48 eV for Si. 
On the other hand, the presented values of $L_{2,3}$ level widths, 0.024 and 0.030 eV for Al and Si respectively, are smaller than the recommended values from Campbell and Papp’s work, i.e. 0.04 eV for Al and 0.05 eV for Si, but they are much larger than values presented in Krause and Oliver’s work, i.e. 0.004 eV for Al and 0.015 eV for Si. 

The presented values of $K$-shell fluorescence yields, 0.042\slash 0.044 and 0.054\slash 0.057 eV for Al and Si respectively (for Coulomb\slash Babushkin gauge), are slightly larger than the ones from Krause’s work \cite{krause-omegi}, i.e. 0.039 for Al and 0.050 for Si, but they differ significantly more from the fitted values recommended by Hubbell et al. \cite{hubbell}, i.e. 0.0397 for Al and 0.043 for Si. 
On the other hand, the presented values of $L_{2,3}$-shell fluorescence yields, 1.49$\times10^{-4}$\slash 1.51$\times10^{-4}$ and 1.78$\times10^{-4}$\slash 1.16$\times10^{-4}$ for Al and Si respectively (for Coulomb\slash Babushkin gauge), are appreciably smaller than the ones from Krause’s work, i.e. 7.5$\times10^{-4}$ for Al and 3.7$\times10^{-4}$ for Si.

\begin{figure}[!htb]
\centering
\includegraphics[width=\columnwidth]{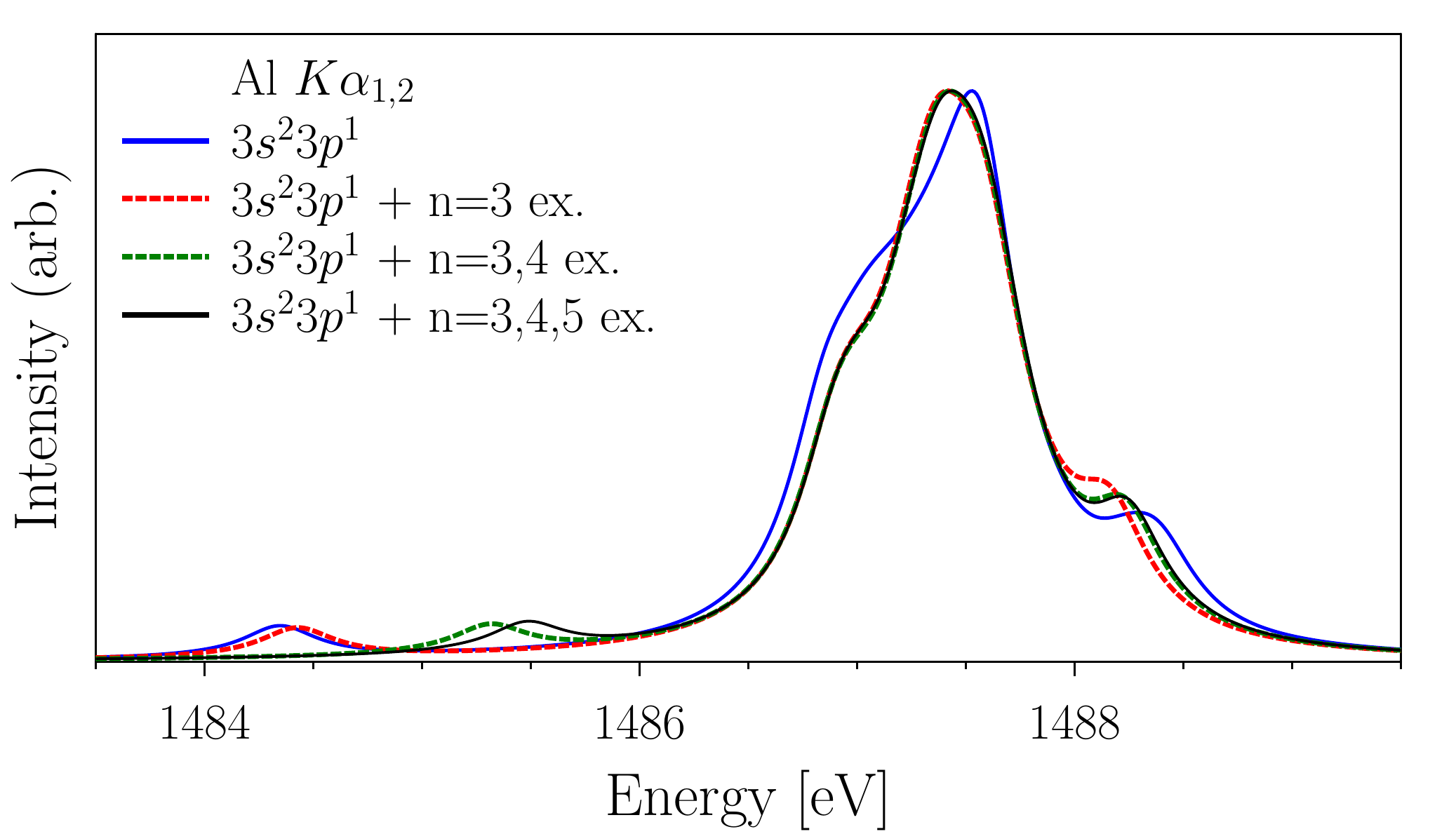}
\caption{\label{fig:al-korel}Simulation of Al $K\alpha_{1,2}$ line shape for various CI basis.}
\end{figure}

\begin{figure}[!htb]
\centering
\includegraphics[width=\columnwidth]{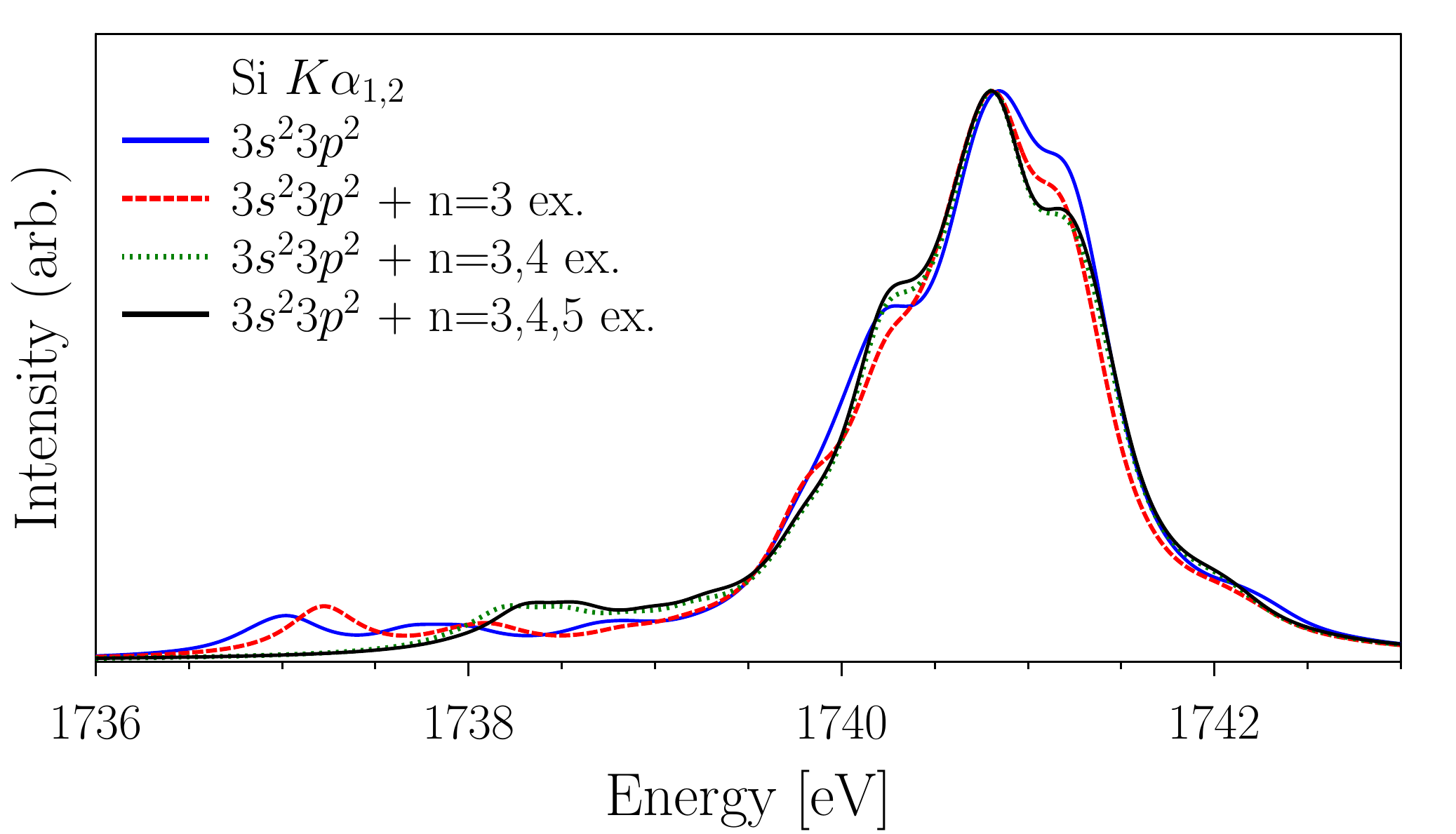}
\caption{\label{fig:si-korel}Simulation of Si $K\alpha_{1,2}$ line shape for various CI basis.}
\end{figure}

\subsection{Width of $K\alpha_{1,2}$ line}

\subsubsection{\label{sec:ovc-rci}OVC predictions}

On Figures \ref{fig:al-korel} and \ref{fig:si-korel} the theoretical simulations of $K\alpha_{1,2}$ line shape for Al and Si have been presented. 
These simulations have been computed on the basis of MCDF-RCI calculations for a sequence of active spaces (in the other words: CSF's bases). 
The ``reference'' basis means CSFs referring to the $3s^23p^1$ and $3s^23p^2$ valence electronic configurations, respectively for Al and Si. 
The ``n=3'' basis means an extended CSF's basis, originated from single (S) and double (D) excitations from the reference $3s^23p^x$ ($x$ = 1, 2) configurations to \{$3s$, $3p$, $3d$\} active space of virtual orbitals. 
Next, ``n=3,4'' and ``n=3,4,5'' bases originate from SD excitations from reference configurations to \{$3s$, $3p$, $3d$, $4s$, $4p$, $4d$, $4f$\} and \{$3s$, $3p$, $3d$, $4s$, $4p$, $4d$, $4f$, $5s$, $5p$, $5d$, $5f$\} active space of virtual orbitals, respectively. 
In the above-mentioned cases the core is \{$1s$, $2s$, $2p$\} orbital space. 

In Tables \ref{tab:al-korel} and \ref{tab:si-korel} the $K\alpha_{1,2}$ x-ray line parameters are collected for Al and Si for various CI basis used in calculations. 
The parameters are: the energy of main peak, $E_{K\alpha_{1,2}}$ (the highest point of synthetic spectrum), and the width of the $K\alpha_{1,2}$ line, $\Gamma_{K\alpha_{1,2}}$, given in two modes: as a width of a main peak fitted with one Lorentz profile (marked ``fit'' in the tables) and as a full width at half maximum (FWHM) of asymmetric peak. 
The linewidths computed according to Eq.~\eqref{eq:ka}, by using recommended values from Campbell and Papp’s work \cite{papp} and from Krause and Oliver’s work \cite{krause} are also presented. 
One can see that despite extending CI basis the values of $E_{K\alpha_{1,2}}$ and $\Gamma_{K\alpha_{1,2}}$ change only slightly. 
The experimental values of $K\alpha_{1,2}$ linewidth are 0.816$\pm$0.005 eV and 0.85$\pm$0.04 eV for Al \cite{citrin} and 1.02$\pm$0.13 eV for Si \cite{graeffe}. 
The comparison of present calculated values against experimental ones allows us to conclude that the established theoretical model, based on the MCDF method and OVC approach, is correct. 
It is worth emphasizing that in the case of Al $K\alpha_{1,2}$ spectra, the extending RCI base from ``reference'' to ``n=3'' or ``n=3,4'' or ``n=3,4,5'' improves the theoretical predictions of linewidth (i.e. makes them nearer to experimental values). 
In the case of Si $K\alpha_{1,2}$ spectra the theoretical predictions of linewidth give results a little larger than experimental ones and it is hard to say that extending the RCI base improves the predictions, so in this case more theoretical efforts and more accurate measurements are needed. 
But even in this case the experimental value is closer to the theoretical predictions based on the OVC model than to the $K\alpha_{1,2}$ transition width evaluated in Eq.~\eqref{eq:ka}. 


\begin{table}[!htb]
\caption{\label{tab:al-korel}Theoretical predictions for $K\alpha_{1,2}$ x-ray line parameters for Al, calculated for various CI basis.}
\medskip
\centering
\begin{tabular*}{\linewidth}{@{\extracolsep{\fill}}l cc c cc}
\hline
&&&&&\\[-1.5ex]
CI basis & \multicolumn{2}{c}{Number of CSF's} & $E_{K\alpha_{1,2}}$ & \multicolumn{2}{c}{$\Gamma_{K\alpha_{1,2}}$ [eV]} \\[0.5ex]
\cline{2-3}\cline{5-6}\\[-1.5ex]
& $1s^{-1}$ & $2p^{-1}$ & [eV]& fit & FWHM\\
&&&&&\\[-1.5ex]\hline
&&&&&\\[-1ex]
reference & 4 & 10 & 1487.53 & 0.97 & 0.98\\
n=3 & 89 & 324 & 1487.42 & 0.85 & 0.90\\
n=3,4 & 664 & 2556 & 1487.43 & 0.86 & 0.90\\
n=3,4,5 & 1831 & 7138 & 1487.44 & 0.86 & 0.90\\[0.5ex]
\hline
&&&&&\\[-1.5ex]
+ sat. & & & & 0.94 & 1.05\\[0.5ex]
\hline
&&&&&\\[-1.5ex]
Eq.~\eqref{eq:ka}:& & & & &\\
\quad Ref.~\cite{papp} & & & & \multicolumn{2}{c}{0.41}\\
\quad Ref.~\cite{krause} & & & & \multicolumn{2}{c}{0.42}\\[0.5ex]
\hline
&&&&&\\[-1.5ex]
exp. \cite{citrin} & & & & \multicolumn{2}{c}{0.816$\pm$0.005}\\
& & & & \multicolumn{2}{c}{0.85$\pm$0.04}\\[0.5ex]
\hline
\end{tabular*}
\end{table}

\begin{table}[!htb]
\caption{\label{tab:si-korel}Theoretical predictions for $K\alpha_{1,2}$ x-ray line parameters for Si, calculated for various CI basis.}
\medskip
\centering
\begin{tabular*}{\linewidth}{@{\extracolsep{\fill}}l cc c cc}
\hline
&&&&&\\[-1.5ex]
CI basis & \multicolumn{2}{c}{Number of CSF's} & $E_{K\alpha_{1,2}}$ & \multicolumn{2}{c}{$\Gamma_{K\alpha_{1,2}}$ [eV]} \\[0.5ex]
\cline{2-3}\cline{5-6}\\[-1.5ex]
& $1s^{-1}$ & $2p^{-1}$ & [eV] & fit & FWHM\\
&&&&&\\[-1.5ex]\hline
&&&&&\\[-1ex]
reference & 8 & 21 & 1740.85 & 1.26 & 1.41\\
n=3 & 230 & 800 & 1740.81 & 1.18 & 1.27\\
n=3,4 & 2299 & 8407 & 1740.81 & 1.28 & 1.35\\
n=3,4,5 & 6636 & 24476 & 1740.80 & 1.30 & 1.37\\[0.5ex]
\hline
&&&&&\\[-1.5ex]
+ sat. & & & & 1.34 & 1.44\\[0.5ex]
\hline
&&&&&\\[-1.5ex]
Eq.~\eqref{eq:ka}:& & & & &\\
\quad Ref.~\cite{papp} & & & & \multicolumn{2}{c}{0.48}\\
\quad Ref.~\cite{krause} & & & & \multicolumn{2}{c}{0.50}\\[0.5ex]
\hline
&&&&&\\[-1.5ex]
exp. \cite{graeffe} & & & & \multicolumn{2}{c}{1.02$\pm$0.13}\\[0.5ex]
\hline
\end{tabular*}
\end{table}

\begin{figure}[!htb]
\begin{center}
\includegraphics[width=\columnwidth]{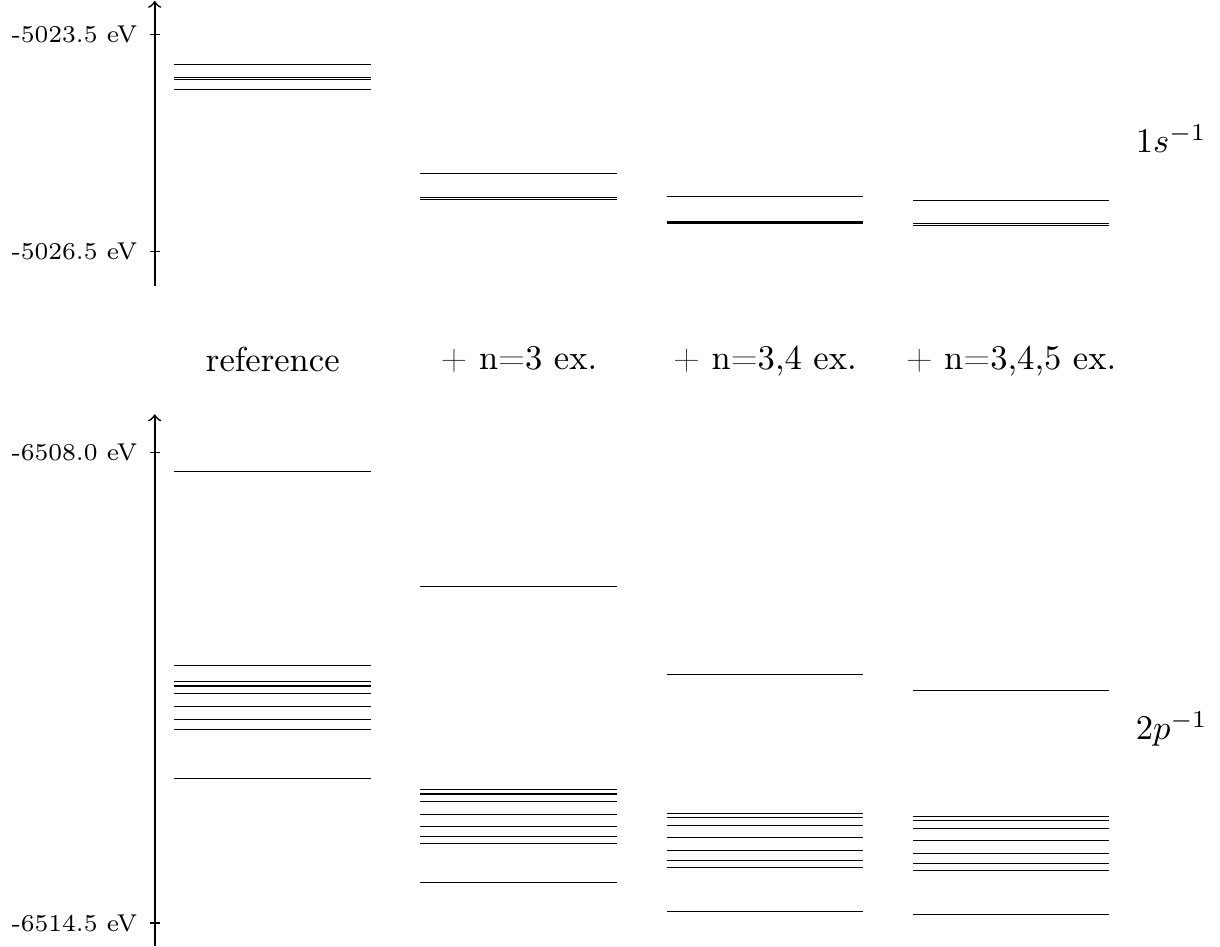}
\end{center}
\caption{Energies of $1s^{-1}$ and $2p^{-1}$ atomic levels for Al for various CI basis. \label{fig:schemat-al-ka-levels}}
\end{figure}

\begin{figure}[!htb]
\begin{center}
\includegraphics[width=\columnwidth]{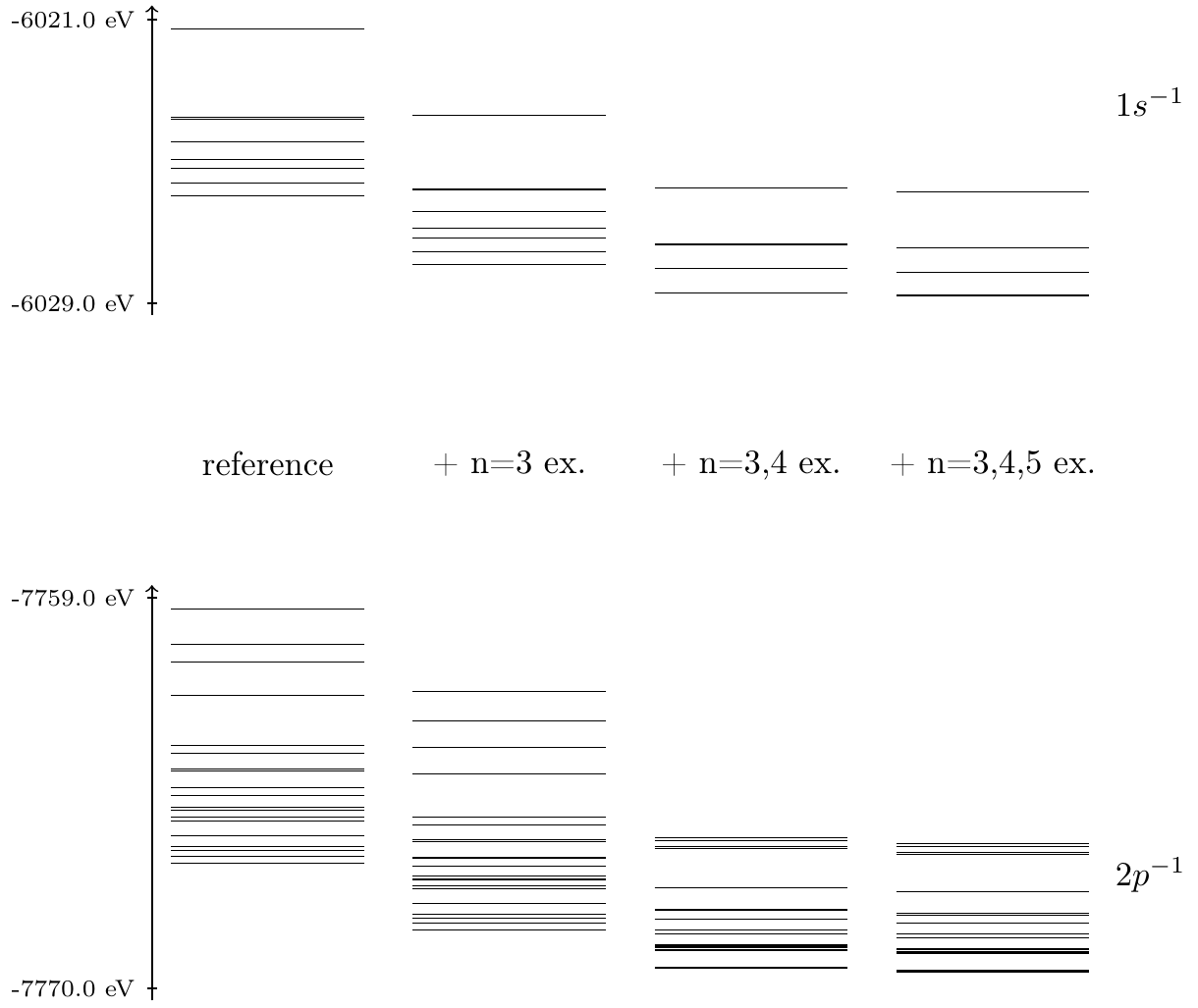}
\end{center}
\caption{Energies of $1s^{-1}$ and $2p^{-1}$ atomic levels for Si for various CI basis. \label{fig:schemat-si-ka-levels}}
\end{figure}

In the Figures \ref{fig:schemat-al-ka-levels} and \ref{fig:schemat-si-ka-levels} the atomic level arrangements of $1s^{-1}$ and $2p^{-1}$ hole levels (i.e. initial and final levels of $K\alpha_{1,2}$ transitions) for Al and Si atoms for various CI bases are presented. 
As one can see, the higher level of electron correlation is included by extending the CI basis of the lower absolute values of $1s^{-1}$ and $2p^{-1}$ hole level energy. 
However, the above mentioned changes of $1s^{-1}$ and $2p^{-1}$ energy levels compensate and the following difference of $K\alpha_{1,2}$ line energy vanishes -- see Tables \ref{tab:al-korel} and \ref{tab:si-korel}, as well as in Figures \ref{fig:al-korel} and \ref{fig:si-korel}.

\subsubsection{\label{sec:satellite}Including satellite contributions}

\begin{figure}[!htb]
\centering
\includegraphics[width=\columnwidth]{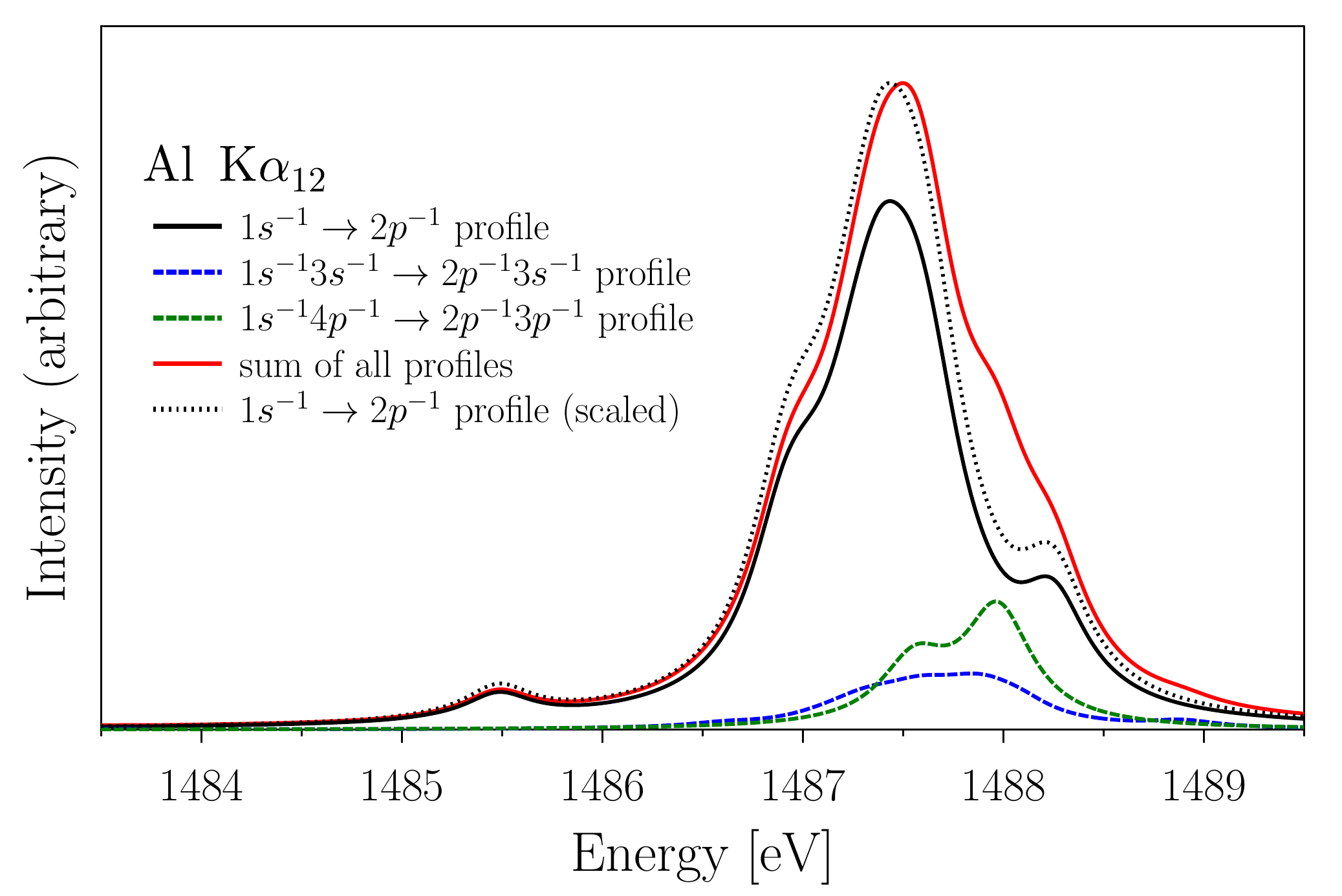}
\caption{\label{fig:al-sat}Simulation of Al $K\alpha_{1,2}$ line shape including satellite contributions originated from $3s$ and $3p$ spectator holes.}
\end{figure}

\begin{figure}[!htb]
\centering
\includegraphics[width=\columnwidth]{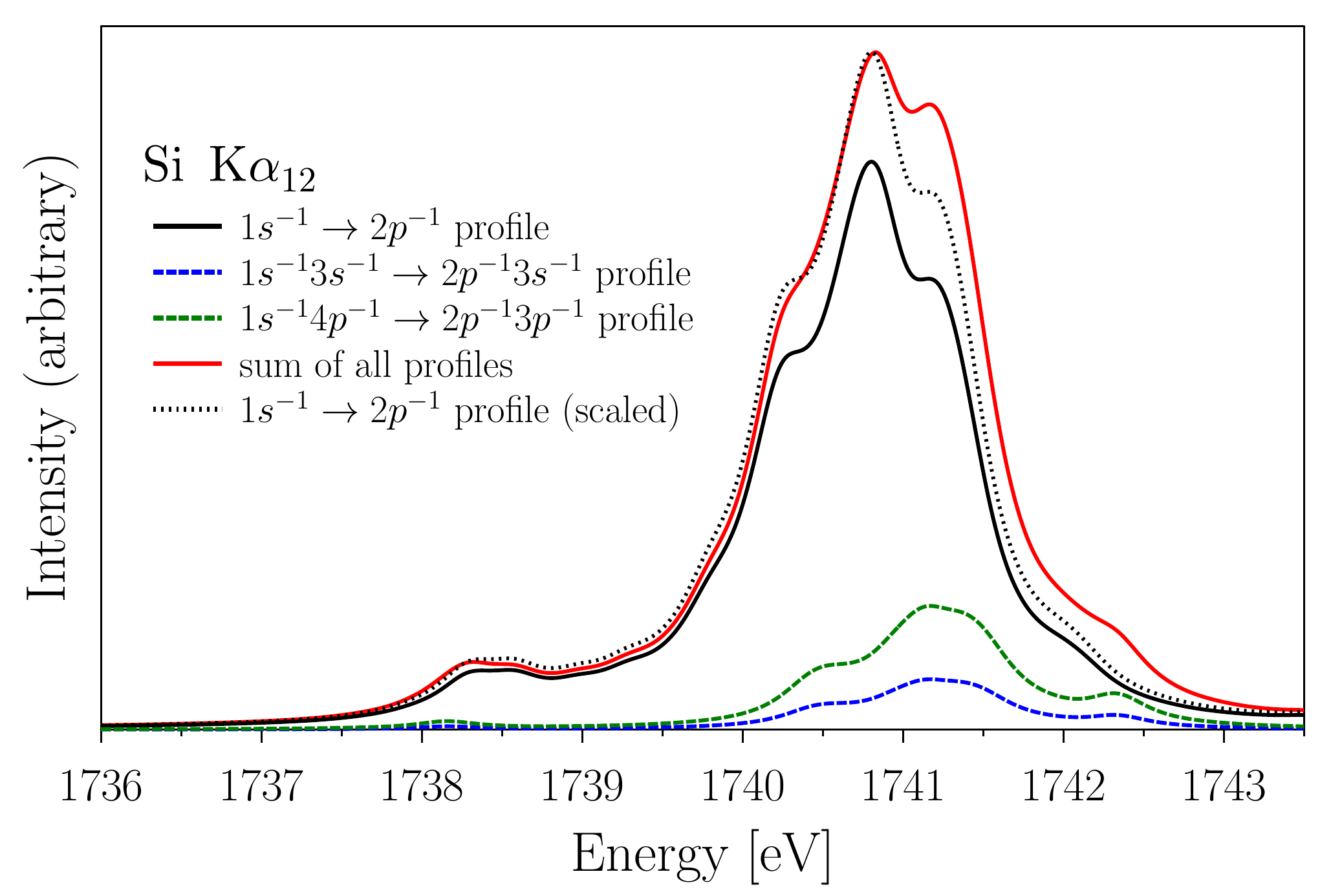}
\caption{\label{fig:si-sat}Simulation of Si $K\alpha_{1,2}$ line shape including satellite contributions originated from $3s$ and $3p$ spectator holes.}
\end{figure}

Another reason for the significant broadening observed experimentally for the x-ray lines can be attributed to the outer-shell ionization and excitation (OIE) effect \cite{lifetimes-prl}. 
In order to evaluate the OIE broadening the calculations of the total shake probabilities, i.e., SO and shakeup (SU), applying the sudden approximation model \cite{carlson} and using MCDF wave functions have been performed. 
The intensity ratio of the intensity of the main $K\alpha_{1,2}$ line, $I_{main}$, to the intensity of the $nl$-shell satellite, $I_{sat}^{nl}$, is given according to binomial distribution: 
\begin{equation}
\frac{I_{sat}^{nl}}{I_{main}}= \frac{N^{nl} \cdot P_{ion}^{nl}}{1-P_{ion}^{nl}}
\end{equation}
where $P_{ion}^{nl}$ is an ionisation probability for the $nl$ shell (due to shake processes) and $N^{nl}$ is a number of electrons occurring on the $nl$ shell. 
The shake probabilities for Al and Si for considered cases of $3s$- and $3p$-hole satellites are presented in Table \ref{tab:shake}. 
In the Figures \ref{fig:al-sat} and \ref{fig:si-sat} the theoretical simulation of Al and Si $K\alpha_{1,2}$ line shapes including satellite contributions originated from the $3s$ and the $3p$ spectator hole have been presented.

\begin{table}[!htb]
\caption{\label{tab:shake}
Total shake probabilities (in percent per subshell) as a result of single $K$-shell ionization calculated for outer shells of Al and Si.}
\medskip
\centering
\begin{tabular*}{\linewidth}{@{\extracolsep{\fill}}l cc}
\hline
&&\\[-1.5ex]
Atom & \multicolumn{2}{c}{Probability [\%] per shell} \\
\cline{2-3}\\[-1.5ex]
& $3s$ & $3p$ \\
&&\\[-1.5ex]\hline
&&\\[-1ex]
Al & 11.81 & 15.08 \\
Si & 7.89 & 18.24 \\[0.5ex]
\hline
\end{tabular*}
\end{table}

As one can see from Tables \ref{tab:al-korel} and \ref{tab:si-korel}, the inclusion of satellite contributions originating from $3s$ and $3p$ spectator holes changes the theoretical predictions for $K\alpha_{1,2}$ linewidth. 
Similar to the results published in Ref.~\cite{lifetimes-prl}, the theoretical synthesised spectrum broadening is observed. 
On the other hand, including satellite contributions in theoretical synthesised spectrum makes it smoother -- see Figures \ref{fig:al-sat} and~\ref{fig:si-sat}.

\subsection{\label{sec:gauges-stab}Controlling convergence in RCI calculations}

The ratio of transition rates calculated by means of Coulomb's (velocity) and Babushkin's (length) gauges can be used as a criterion of calculated wavefunction quality. 
In particular, this approach is helpful in controlling of the convergence in RCI calculations utilising active sets of virtual orbitals. 

The strongest transition within $K\alpha_{1,2}$ line for Al is a transition from   
$$
\ket{i} = a\cdot \ket{1s^1 2s^2 2p_{1/2}^2 2p_{3/2}^4 3s^2 3p_{1/2}^1}_{J=0} + \ldots
$$
initial state to 
$$
\begin{array}{ll}
\ket{f} = &
\phantom{+} b\cdot \ket{1s^2 2s^2 2p_{1/2}^2 2p_{3/2}^3 3s^2 3p_{1/2}^1}_{J=1} \\[2ex]
&+ c\cdot \ket{1s^2 2s^2 2p_{1/2}^1 2p_{3/2}^4 3s^2 3p_{1/2}^1}_{J=1}\\[2ex] 
&+ d\cdot \ket{1s^2 2s^2 2p_{1/2}^2 2p_{3/2}^3 3s^2 3p_{3/2}^1}_{J=1} + \ldots 
\end{array}
$$
final state, where $a$, $b$, $c$, and $d$ are the CSF's mixing coefficients of the largest contribution in the ASF. 
In Table~\ref{tab:gauges-diff} the following parameters, resulting from RCI calculations utilising various CI bases, have been collected: mixing coefficients for initial state ($a$) and for final state ($b$, $c$, and $d$), the transition rate for mentioned transition calculated by means of Coulomb's ($I_C$) and Babushkin's ($I_B$) gauges, and the Babushkin's\slash{}Coulomb's rate ratios ($I_B/I_C$).

\begin{table*}[htb!]
\caption{\label{tab:gauges-diff}The mixing coefficients for initial state and final state of the strongest $K\alpha_{1,2}$ transition for Al, the transition rate calculated by Coulomb and Babushkin gauges, and the Babushkin/Coulomb rate ratios (see text for details).}
\begin{tabular*}{\linewidth}{@{\extracolsep{\fill}}l cccc cc c}
\hline
\multicolumn{8}{c}{~}\\[-1.5ex]
CI basis & $a$ & $b$ & $c$ & $d$ & $I_C$~[$\times10^{13}$~s$^{-1}$] & $I_B$~[$\times10^{13}$~s$^{-1}$] & $I_B/I_C\cdot100\%$\\[0.7ex]
\hline\\[-1ex]
reference	& 1 & 0.3559 & 0.3272 & 0.2847 & 1.6172 & 1.7333 & 107.178\\[0.5ex]
n=3	& 0.9301 & 0.3436 & 0.2812 & 0.2769 & 1.6008 & 1.7153 & 107.151\\[0.5ex]
n=3,4	& 0.9279 & 0.3338 & 0.2938 & 0.2688 & 1.6050 & 1.7198 & 107.153\\[0.5ex]
n=3,4,5	& 0.9284 & 0.3350 & 0.2894 & 0.2714 & 1.5992 & 1.7136 & 107.154\\[1ex]
\hline
\end{tabular*}
\end{table*}

As one can see from the data collected in Table~\ref{tab:gauges-diff}, despite extending the CSF's basis, the calculation numeric stability is preserved because the $I_B/I_C$ ratio changes only minimally. 
The difference of transition rates calculated by means of Coulomb's and Babushkin's gauges is only about 7\%. 
If the increasing of CSF's basis size would increase the $I_B/I_C$ ratio, it would mean that the wavefunctions determined for a larger CSF's basis (generated by using a larger number of virtual orbitals) are of lower quality than those determined for a smaller base. 
In the present work the increasing CSF's basis size does not lead to a significant increase in the $I_B/I_C$ ratio -- this indicates that the quality of wavefunction for virtual spinorbitals is not worse than that for real ones.

\subsection{Shape of $K\alpha_{1,2}$ line}

\begin{figure}[!htb]
\centering
\includegraphics[width=\columnwidth]{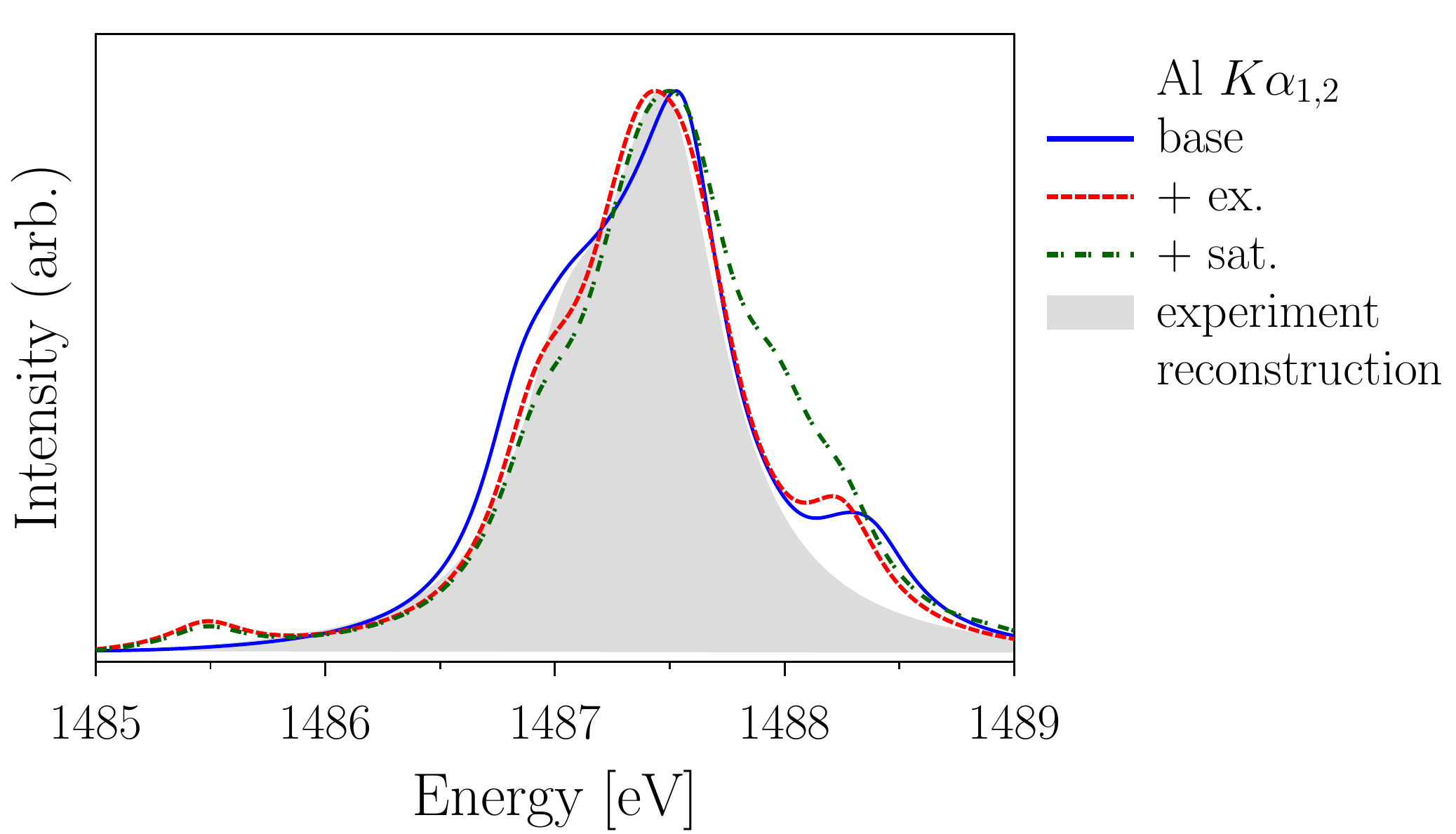}
\caption{\label{fig:al-exp-reco}Simulations of Al $K\alpha_{1,2}$ line shape compared to experiment reconstruction.}
\end{figure}

\begin{figure}[!htb]
\centering
\includegraphics[width=\columnwidth]{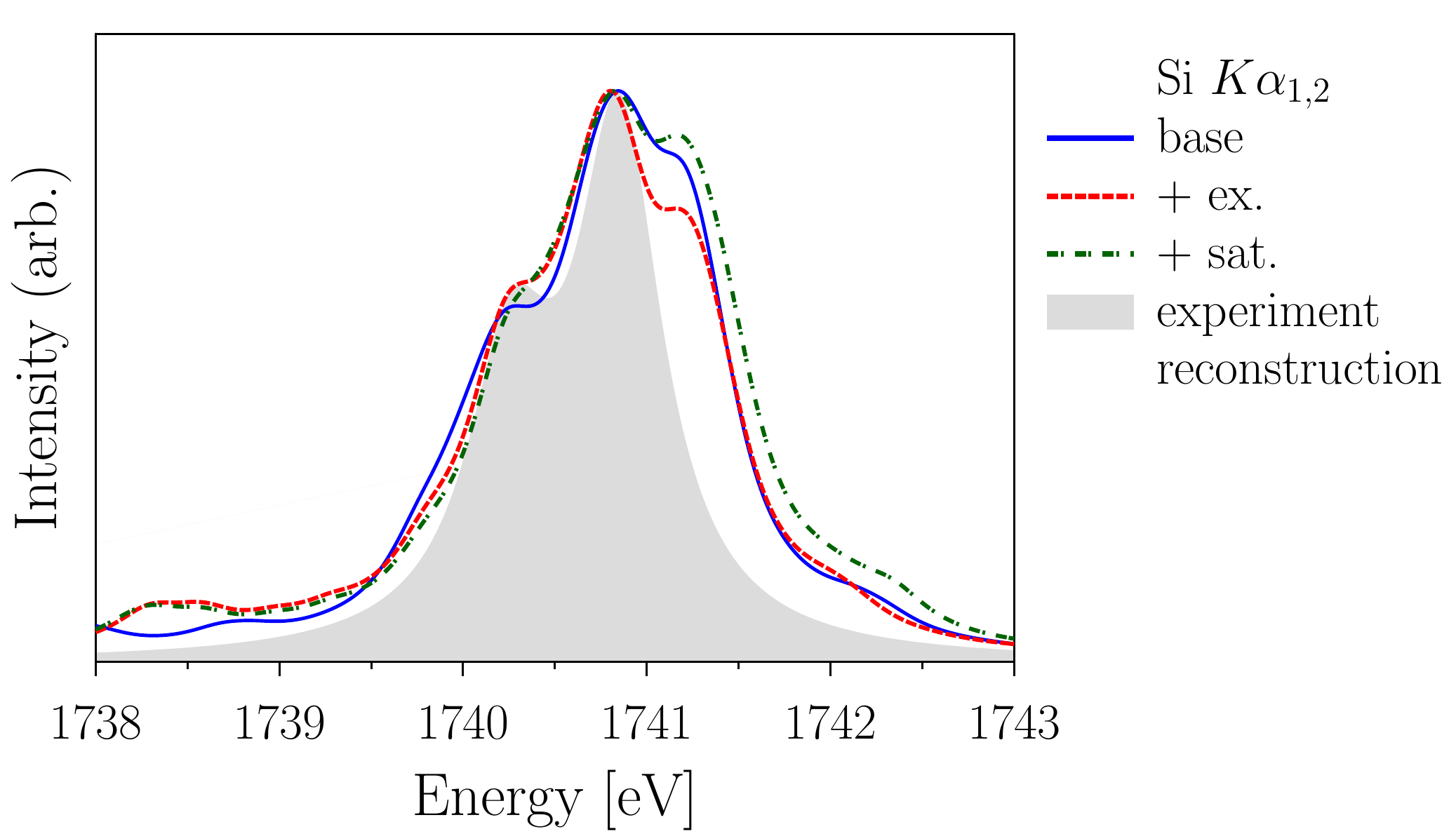}
\caption{\label{fig:si-exp-reco}Simulations of Si $K\alpha_{1,2}$ line shape compared to experiment reconstruction.}
\end{figure}

In Figs.~\ref{fig:al-exp-reco} and~\ref{fig:si-exp-reco} the experimental $K\alpha_{1,2}$ line shapes for Al and Si have been compared to the theoretical predictions. 
Experimental shapes have been reconstructed from fitting parameters (width of Lorentz profiles assigned to $K\alpha_{1}$ and $K\alpha_{2}$ lines, intensity ratio of these Lorentzians) presented in Ref.~\cite{al-decomposition} and Ref.~\cite{graeffe} for Al and Si, respectively. 
Three kinds of theoretical simulations are presented in Figs.~\ref{fig:al-exp-reco} and~\ref{fig:si-exp-reco}: ``base'' -- calculation for $1s^{-1} \to 2p^{-1}$ transitions by using ``reference'' basis set; ``+ex.'' -- calculation for $1s^{-1} \to 2p^{-1}$ transitions by using the basis set extended by SD excitations to $n$=3,4,5 virtual orbitals (see section~\ref{sec:ovc-rci} for details); ``+sat.'' -- shape, also including satellite contributions (see section \ref{sec:satellite} for details).

In the case of Al the theoretical simulated shapes of the $K\alpha_{1,2}$ line without satellite contributions are in close agreement with the experiment reconstruction. The main difference is a bulge appeared in the theoretical simulation on the high-energy side. 
Adding satellite contributions increases the difference between theoretical and experimental shapes. 
This can indicate that in experimental circumstances of Ref.~\cite{al-decomposition} there is no sudden change in condition  \cite{carlson} and no shake-originated satellite contributions are created. 

In the case of Si experimental reconstructed $K\alpha_{1,2}$ line differs significantly from all three theoretical simulations on the high-energy side, although they fit well on the low-energy side. 
This difference can be attributed to the influence of the chemical environment that changes the arrangement of the energy levels assigned to $1s^{-1}$ and $2p^{-1}$ hole states over those predicted by the theory of isolated atoms \cite{hartmann1,hartmann2}. 
It is worth noting that although Al and Si are neighbours in the periodic table of elements, they have significantly different chemical environment. The most important point is that the Al atoms are bonded within a solid by metallic bonds, but the Si atoms -- by covalent bonds. 
So, a proper explanation of the structure of x-ray lines for low-Z elements, for which the chemical environment can noticeably influence the hole states arrangement, requires a more complex molecular description and computation.

\section{Conclusions}

In this work MCDF and RCI methods have been employed to predict the structure and the width of $K\alpha_{1,2}$ x-ray lines of Al and Si, followed by computation of $K$ and $L_{2,3}$ atomic level widths. 
The influence of electron correlation on $1s^{-1}$ and $2p^{-1}$ hole state levels energy and the $K\alpha_{1,2}$ spectra structure has been also studied. 

Based on the results presented above, some general conclusions can be drawn:
(a) It is possible to make theoretical \textit{ab initio} predictions for $K\alpha_{1,2}$ linewidths, basing on the OVC model, which are in agreement with experimental data; 
(b) There are more extensive electron correlation inclusion influences visible on hole state energy levels, but only a little on $K\alpha_{1,2}$ line energy and effective width, and\ldots{} 
(c) The case of the Al $K\alpha_{1,2}$ line shows that more extensive electron correlation inclusion improves the predictions of $K\alpha_{1,2}$ linewidth; in the case of the Si $K\alpha_{1,2}$ line, the effect of electron correlation inclusion is not clear;
(d) Including satellite contributions originated from $3s$ and $3p$ spectator holes in theoretical synthesized Al and Si $K\alpha_{1,2}$ spectra results in spectra broadening, but unfortunately it increases the difference between theory vs. experiment. 
Thus, more theoretical effort is needed to provide a proper explanation of Al and Si $K\alpha_{1,2}$ lines’ width and structure. 
On the other hand, new high-resolution Al and Si $K\alpha_{1,2}$ x-ray spectra measurements, to prove or reject theoretical predictions, are very welcome.

%

\end{document}